\documentclass[aps,pra,showpacs, twocolumn]{revtex4}
\usepackage{graphicx}    
\usepackage{amsmath}

\begin{document}

\title{Nonclassicality criteria: Quasiprobability distributions and correlation functions}

\author{Moorad Alexanian}
\email[]{alexanian@uncw.edu}

\affiliation{Department of Physics and Physical Oceanography\\
University of North Carolina Wilmington\\ Wilmington, NC
28403-5606\\}

\date{\today}

\begin{abstract}
We use the exact calculation of the quantum mechanical, temporal characteristic function $\chi(\eta)$ and the degree of second-order coherence $g^{(2)}(\tau)$ for a single-mode, degenerate parametric amplifier for a system in the Gaussian state, viz., a displaced-squeezed thermal state, to study the different criteria for nonclassicality. In particular, we contrast criteria that involve only one-time functions of the dynamical system, for instance, the quasiprobability distribution $P(\beta)$ of the Glauber-Sudarshan coherent or P-representation of the density of state and the Mandel $Q_{M}(\tau)$ parameter, versus the criteria associated with the two-time correlation function $g^{(2)}(\tau)$.
\end{abstract}


\maketitle {}
\section{Introduction}

The field of quantum computation and quantum information, as applied to quantum computers, quantum cryptography, and quantum teleportation, was originally based on the manipulation of quantum information in the form of discrete quantities like qubits, qutrits, and higher-dimensional qudits. Nowadays the emphasis has shifted on processing quantum information by the use of continuous-variable quantum information carriers. In this regard, use is now made of any combination of Gaussian states, Gaussian operations, and Gaussian measurements \cite{WPP12, ARL14}. The interest in Gaussian states is both theoretical as well as experimental since simple analytical tools are available and, on the experimental side, optical components effecting Gaussian processes are readily available in the laboratory \cite{WPP12}.

Quantum optical systems give rise to interesting nonclassical behavior such as photon antibunching and sub-Poissonian photon statistics owing to the discreetness or photon nature of the radiation field \cite{GSA13}. These nonclassical features can also be quantified with the aid of the temporal second-order quantum mechanical correlation function $g^{(2)}(\tau)$ and experimentally studied using a Hanbury Brown--Twiss intensity interferometer modified for homodyne detection \cite{GSSRL07}. Physical realizations and measurements of the second-order coherence function $g^{(2)}(\tau)$ of light have been studied earlier via a degenerate parametric amplifier (DPA) \cite{KHM93,LO02,GSSRL07}.

The early work on parametric amplification \cite{MGa67, MGb67} has led to a wealth of research, for instance, in sub-Poissonian statistics and squeezed light \cite{AZ92}, squeezing in the output of a cavity field \cite{BLPa90, BLPb90}, quantum noise, measurement, and amplification \cite{CDG10}, and photon antibunching \cite{HP82}.

The need to formulate measurable conditions to discern the classical or nonclassical behavior of a dynamical system is important and so several criteria exist for nonclassicality. In particular, the use of the Glauber-Sudarshan P function to determine the existence or nonexistence of a quasiprobability distribution $P(\beta)$ that would characterize whether the system has a classical counterpart or not \cite {RSA15}. The existent differing criteria for nonclassicality actually complement each other since nonclassicality criteria derived from the one-time function $P(\beta)$ is actually complemented by the nonclassicality criteria involving the two-time coherence function $g^{(2)}(\tau)$. Note that nonclassicality information provide by $g^{(2)}(\tau)$ cannot be obtained solely from $P(\beta)$.

In a recent work \cite{MA16}, a detailed study was made of the temporal development of the second-order coherence function for Gaussian states---displaced-squeezed thermal states---the dynamics being governed by a Hamiltonian for degenerate parametric amplification. The time development of the Gaussian state is generated by an initial thermal state and the system subsequently evolves in time where the usual assumption of statistically stationary fields is not made \cite{SZ97}.

In the present work, we compare the differing criteria for nonclassicality. In Section II, we consider the general Hamiltonian of the degenerate parametric amplifier. In Section III, we find an exact expression for the characteristic function and introduce the Glauber-Sudarshan coherent state or P-representation of the density matrix. In Section IV, we obtain, via a two-dimensional Fourier transform, the quasiprobability distribution $P(\beta)$ from the exact expression of the characteristic function $\chi(\eta)$ and obtain from $P(\beta)$ the necessary and sufficient condition for nonclassicality.  In Section V, we present the known nonclassicality criteria for the coherence function $g^{(2)}(\tau)$. In Section VI, we study numerical examples to elucidate how all the different criteria for nonclassicality complement each other. Finally, Section VII summarizes our results.

\section {Degenerate parametric amplification}

The Hamiltonian for degenerate parametric amplification, in the interaction picture, is
\begin{equation}
\hat{H} = c \hat{a}^{\dag 2} + c^* \hat{a}^2 + b\hat{a} + b^* \hat{a}^\dag.
\end{equation}
The system is initially in a thermal state $\hat{\rho}_{0}$ and a after a preparation time $t$, the system temporally develops into a Gaussian state and so \cite{MA16}
\begin{equation}
\hat{\rho}_{G}=\exp{(-i\hat{H}t/\hbar)}\hat{\rho}_{0} \exp{(i\hat{H}t/\hbar)}
\end{equation}
\[
= \hat{D}(\alpha) \hat{S}(\xi)\hat{\rho}_{0} \hat{S}(-\xi) \hat{D}(-\alpha),
\]
with  the displacement $\hat{D}(\alpha)= \exp{(\alpha \hat{a}^{\dag} -\alpha^* \hat{a})}$  and the squeezing $\hat{S}(\xi)=  \exp\big{(}-\frac{\xi}{2} \hat{a}^{\dag 2} + \frac{\xi^*}{2} \hat{a}^{2} \big{ )}$ operators, where $\hat{a}$ ($\hat{a}^{\dag})$ is the photon annihilation (creation) operator, $\xi = r \exp{(i\theta)}$, and $\alpha= |\alpha|\exp{(i\varphi)}$. The thermal state is given by

\begin{equation}
\hat{\rho}_{0} = \exp{(- \hbar \omega\hat{n}/k_{B}T)}/ \textup{Tr}[\exp{(- \hbar \omega \hat{n}/k_{B}T)}],
\end{equation}
with $\hat{n}= \hat{a}^\dag \hat{a}$ and $\bar{n}= \textup{Tr}[\hat{\rho}_{0} \hat{n}]$ .

The parameters $c$ and $b$ in the degenerate parametric Hamiltonian (1) are determined \cite{MA16} by the parameters  $\alpha$ and $\xi$ of the Gaussian density of state (2) via
\begin{equation}
tc = -i\frac{\hbar}{2} r\exp(i\theta)
\end{equation}
and
\begin{equation}
tb= -i\frac{\hbar}{2}\Big{(} \alpha \exp{(-i\theta)} + \alpha^* \coth (r/2)\Big{)} r,
\end{equation}
where $t$ is the time that it takes for the system governed by the Hamiltonian (1) to generate the Gaussian density of state $\hat{\rho}_{G}$ from the initial thermal density of state $\hat{\rho}_{0}$.

The quantum mechanical seconde-order, degree of coherence is given by \cite{MA16}
\begin{equation}
g^{(2)}(\tau) = \frac{\langle \hat{a}^{\dag}(0) \hat{a}^{\dag}(\tau) \hat{a}(\tau) \hat{a}(0)\rangle }{\langle \hat{a}^{\dag}(0)  \hat{a}(0)\rangle \langle \hat{a}^{\dag}(\tau)\hat{a}(\tau) \rangle},
\end{equation}
where all the expectation values are traces with the Gaussian density operator, viz., a displaced-squeezed thermal state. Accordingly, the system is initially in the thermal state $\hat{\rho}_{0}$. After time $t$, the system evolves to the Gaussian state $\hat{\rho}_{G}$ and a photon is annihilated at time $t$, the system then develops in time and after a time $\tau$ another photon is annihilated \cite{MA16}. Therefore, two photon are annihilated in a time separation $\tau$ when the system is in the Gaussian density state $\hat{\rho}_{G}$.

It is important to remark that we do not suppose statistically stationary fields \cite{SZ97}. Therefore, owing to the $\tau$ dependence of the number of photons in the cavity in the denominator of Equation (6), the system asymptotically, as $\tau\rightarrow \infty$, approaches a finite limit without supposing any sort of dissipative processes \cite{MA16}. The coherence function $g^{(2)}(\tau)$ is a function of $\Omega \tau=(r/t)\tau$, $\alpha$, $\xi$, and the average number of photons $\bar{n}$ in the initial thermal state (3), where the preparation time $t$ is the time that it takes the system to dynamically generate the Gaussian density $\hat{\rho}_{G}$ given by (2) from the initial thermal state $\hat{\rho}_{0}$ given by (3). Note that the limit $r\rightarrow 0$ is a combined limit whereby $\Omega =r/t$ also approaches zero resulting in a correlation function which has a power law decay as $\tau/t \rightarrow \infty$ rather than an exponential law decay as $\tau/t \rightarrow \infty$ as is the case in the presence of squeezing when $r>0$ \cite{MA16}.

\section{characteristic function}

The calculation of the correlation function (6) deals with the product of two-time operators. However, a complete statistical description of a field involves only the expectation value of any function of the operators $\hat{a}$ and $\hat{a}^\dag$. A characteristic function contains all the necessary information to reconstruct the density matrix for the state of the field.

Now \cite{MA16}
\begin{equation}
\hat{\rho}(t+\tau) =\exp\big{(}-i\hat{H}(t+\tau)\big{)} \hat{\rho}_{0}\exp \big{(}i\hat{H}(t+\tau)\big{)}
\end{equation}
\[
= \exp(-i\hat{H}\tau) \hat{\rho}_{G}\exp(i\hat{H}\tau).
\]
Accordingly, for any operator function $\mathcal{\hat{O}}(\hat{a},\hat{a}^\dag)$, one has that
\begin{equation}
\textup{Tr}[\hat{\rho}(t+\tau) \mathcal{\hat{O}}(\hat{a},\hat{a}^\dag)] = \textup{Tr}[\hat{\rho}_{G} \mathcal{\hat{O}}\big{(}\hat{a}(\tau),\hat{a}^\dag (\tau)\big{)}]
\end{equation}
\[
\equiv \langle  \mathcal{\hat{O}}\big{(}\hat{a}(\tau),\hat{a}^\dag (\tau)\big{)} \rangle .
\]

One obtains for the characteristic function
\[
\chi(\eta) = \textup{Tr}[\hat{\rho}(t+\tau)\exp{(\eta \hat{a}^\dag}-\eta^*\hat{a})]\exp{(|\eta|^2/2)}
\]
\[
=\textup{Tr}[\hat{\rho}(t+\tau)\exp{(\eta \hat{a}^\dag}) \exp{(-\eta^*\hat{a})}]
\]
\begin{equation}
=\exp{(|\eta|^2/2)} \exp{\big{(}\eta A^*(\tau)- \eta^* A(\tau)\big{)}}\cdot
\end{equation}
\[
\cdot \exp{\big{(}-(\bar{n}+1/2)|\xi(\tau)|^2\big{)}},
\]
where

\[
A(\tau) =\alpha \Bigg{(}\cosh(\Omega\tau)+\frac{1}{2}\coth(r/2) \sinh (\Omega \tau)
\]
\begin{equation}
-\frac{1}{2} (\cosh(\Omega \tau)-1)+\exp[i(\theta -2 \varphi)]\Big{[} -\frac{1}{2}\sinh(\Omega\tau)
\end{equation}
\[
-\frac{1}{2}\coth(r/2)\big{(}\cosh(\Omega\tau)-1\big{)}\Big{]} \Bigg{)}
\]
and
\begin{equation}
\xi(\tau)= \eta\cosh(\Omega \tau +r) +\eta^* \exp(i\theta) \sinh(\Omega \tau +r).
\end{equation}

The expectation value $\textup{Tr} [\hat{\rho}(t+\tau)\hat{a}^{\dag m}\hat{a}^n]$ can be calculated by differentiation of the characteristic function $\chi(\eta)$ with respect to $\eta$ and $\eta^*$ as independent variables, viz., $\textup{Tr} [\hat{\rho}(t+\tau)\hat{a}^{\dag m}\hat{a}^n]= (\partial/\partial \eta)^m (-\partial/\partial \eta^*)^n  \chi(\eta)\Big{|}_{\eta=0}$. Accordingly, knowledge only of the characteristic function can determine only one-time properties of the dynamical system.

Define
\begin{equation}
|\xi(\tau)|^2 = \eta^2 T^*(\tau) +\eta^{*2} T(\tau) +\eta \eta^* S(\tau),
\end{equation}
with
\begin{equation}
T(\tau)= \frac{1}{2} \exp{(i \theta)} \sinh [2(\Omega \tau + r)]
\end{equation}
and
\begin{equation}
S(\tau)= \cosh[2(\Omega \tau +r)].
\end{equation}

In the Glauber-Sudarshan coherent state or P-representation of the density operator $\hat{\rho}$ one has that \cite{GSA13}
\begin{equation}
\hat{\rho} =\int \textup{d}^2\beta \hspace{0.05in} P(\beta)|\beta\rangle \langle\beta| ,
\end{equation}
where $|\beta\rangle$ is a coherent state and nonclassicality occurs when $P(\beta)$ takes on negative values and becomes more singular than a Dirac delta function. One has the normalization condition $\int P(\beta) \textup{d}^2\beta=1$; however, $P(\beta)$ would not describe probabilities, even if positive, of mutually exclusive states since coherent states are not orthogonal. In fact, coherent states are over complete.

The quasiprobability distribution $P(\beta)$ is related to the characteristic function $\chi(\eta)$ via the two-dimensional Fourier transform
\begin{equation}
P(\beta)= \frac{1}{\pi^2} \int \textup{d}^2\eta \hspace{0.05in}\chi(\eta) \exp{(-\beta^*\eta +\beta \eta^*)}.
\end{equation}
The characteristic function $\chi(\eta)$ is a well-behaved function whereas the integral (16) is not always well-behaved, for instance, if $\chi(\eta)$ diverges as $|\eta|\rightarrow \infty$, then $P(\beta)$ can only be expressed in terms of generalized functions. Nonetheless, $P(\beta)$ can be still used to calculate moments of products of $\hat{a}$ and $\hat{a}^\dag$.

It is important to remark that knowledge of $P(\beta)$ without further knowledge of the dynamics governing the system, can only be used to calculate equal-time properties of the system and does not allow us to calculate, for instance, correlation functions, in particular, the quantum mechanical, second-order degrees of coherence $g^{(2)}(\tau)$. The determination of the latter requires, in addition, to  $P(\beta)$ the temporal behavior $\hat{a}(\tau)$.

\section{P-representation}

The integral (16) can be carried out for the characteristic function (9) and so
\begin{equation}
P(\beta)=\frac{2}{\pi} \frac{1}{\sqrt{4a^2b^2-c^2}} e^{-(a^2f^2+b^2d^2+cfd)/(4a^2b^2-c^2)},
\end{equation}
where
\[
a^2= -\frac{1}{2} +(\bar{n}+1/2)\big{(}T(\tau) +T^*(\tau) +S(\tau)\big{)},
\]
\[
b^2= -\frac{1}{2} -(\bar{n}+1/2)\big{(}T(\tau) +T^*(\tau) -S(\tau)\big{)},
\]
\begin{equation}
c=-2 i(\bar{n}+1/2)\big{(}T^*(\tau) -T(\tau)\big{)},
\end{equation}
\[
d= i(A(\tau) -A^*(\tau)-\beta+ \beta^*),
\]
\[
f= A(\tau)+A^*(\tau) -\beta-\beta^*.
\]

The existence of a real-valued function $P(\beta)$ requires that $(4a^2b^2-c^2)\geq 0$, which with the aid of (18), gives that
\begin{equation}
1\leq (2\bar{n} +1) e^{-2(\Omega \tau+r)},
\end{equation}
where the equality hold when $\bar{n}=0$ and $r=0$ at $\tau=0$. Note that criterion (19) does not depend on the coherent amplitude $\alpha$, which appears via $A(\tau)$. Also, if inequality (19) is initially satisfied at $\tau=0$, then as time goes on the inequality will be violated since the squeezing continues indefinitely and so no matter the value of $\bar{n}$, eventually as $\tau$ increases the dynamics will always lead to nonclassical states.

The existence of $P(\beta)$ requires also that it must vanish as $|\beta|\rightarrow \infty$. The bilinear form $(a^2f^2+b^2d^2+cfd)$ in the exponential in (17) can be diagonalized in the variables $\Re{(A(\tau)-\beta)}$  and $\Im{(A(\tau)-\beta)}$ resulting in the eigenvalues $2\big{(}-1+(2\bar{n} +1)e^{2(\Omega \tau+r)}\big{)}$ and $2\big{(}-1+(2\bar{n} +1)e^{-2(\Omega \tau+r)}\big{)}$ that must be nonnegative which requirement gives rise to the same condition (19) for the existence of a genuine probability distribution $P(\beta)$.

Two simple examples follow directly from (16). For the displaced vacuum state for $\tau\geq 0$, one obtains, since $\Omega =r/t =0$, that $P(\beta)=\delta(\beta-\alpha)$, the coherent state. Similarly, for $\bar{n} > 0$ one obtains for the displaced thermal state that $P(\beta)= (1/(\pi \bar{n}))\exp {(-|\beta-\alpha|^2/\bar{n})}$, which becomes the previous example in the vacuum limit when $\bar{n}\rightarrow 0$.

The necessary and sufficient condition for nonclassicality is then
\begin{equation}
(2\bar{n} +1) e^{-2(\Omega \tau+r)} <1,
\end{equation}
which is based only on knowledge of $P(\beta)$. Note that (20) is independent of the value of the coherent parameter $\alpha$.

\section{Nonclassicality criteria}

As indicated above, mere knowledge of $P(\beta)$ does not allow the calculation of the quantum mechanical correlation functions additional knowledge of the the dynamics of the system is necessary, for instance, $\hat{a}(\tau)$ for $\tau\geq 0$. Nonclassical light can be characterized differently, for instance, with the aid of the quantum degree of second-order coherence $g^{(2)}(\tau)$ by the nonclassical inequalities
\begin{equation}
g^{(2)}(0)< 1 \hspace{0.3in} \textup{and}    \hspace{0.3in} g^{(2)}(0) <  g^{(2)}(\tau),
\end{equation}
where the first inequality represents the sub-Poissonian statistics, or photon-number squeezing, while the second gives rise to antibunched light. Hence a measurement of $g^{(2)}(\tau)$ can be used to determine the nonclassicality of the field. The two nonclassical effects often occur together but each can occur in the absence of the other. Similarly, one can derive the nonclassical inequality \cite{RC88}
\begin{equation}
|g^{(2)}(0)-1| < |g^{(2)}(\tau)-1|,
\end{equation}
that is, $g^{(2)}(\tau)$ can be farther away from unity than it was initially at $\tau=0$.

Accordingly, in the determination of the nonclassicality of the field, situations may arise where some of the observable nonclassical characteristics such as squeezing and sub-Poissonian statistics are lost while $P(\alpha)$ still remains nonclassical, that is, inequality (20) holds true while some of the inequalities in (21) and (22) are violated. These situations do arise since the nonclassicality condition (20) is independent of the value of the coherent amplitude $\alpha$ whereas the nonclassicality conditions (21) and (22) do depend on the value of $\alpha$.

Another sufficient condition for nonclassicality is determined by the Mandel $Q_{M}(\tau)$ parameter related to the photon-number variance \cite{GSA13, SZ97}
\begin{equation}
Q_{M}(\tau)= \frac{\Delta n^2(\tau) -\langle \hat{n}(\tau)\rangle}{\langle \hat{n}(\tau)\rangle},
\end{equation}
where $-1 \leq Q_{M}(\tau) <0$ implies that $P(\alpha)$ assumes negative values and thus the field must be nonclassical with sub-Poissonian statistics. Condition $Q_{M}(0) <0$ is equivalent to the first condition in Equation (21) since $Q_{M}(0) = \langle \bar{n}(0)\rangle[g^{(2)}(0)-1]$. It important to remark that the latter equality holds only at $\tau=0$ when both $Q_{M}(0)$ and $g^{(2)}(0)$ represent one-time functions. The correlation function $g^{(2)}(\tau)$ is a two-time function for $\tau >0$ whereas $Q_{M}(\tau)$ is a one-time function for $\tau \geq 0$.  Note  that if the Mandel $Q_{M}(\tau)$ parameter is positive, then no conclusion can be drawn on the nonclassical nature of the radiation field.

The evaluation of $G_{M}(\tau)$ requires knowledge of the characteristic function $\chi(\eta)$ or the quasiprobability distribution $P(\beta)$ and by taking successive derivatives. Such knowledge involves only one-time functions; whereas the correlation function $g^{(2)}(\tau)$ is a two-time function thus the nonclassicality determined by differing criteria complement each other.

\section{Numerical comparisons}

Owing to the equivalence of the nonclassical conditions given by the first of Equation (21) and the Mandel condition $Q_{M}(0)<0$, we need study only numerically the relation of the nonclassical inequalities (21) and (22) for the coherence function $g^{(2)}(\tau)$ and compare them to the nonclassical condition (20) for the quasiprobability distribution $P(\beta)$. It is important to remark that the nonclassicality criteria (21) and (22) for $g^{(2)}(\tau)$ depend strongly of the value of the coherent amplitude $\alpha$ whereas the nonclassicality criterion (20) for $P(\alpha)$ is actually independent of the value of $\alpha$. The coherence function $g^{(2)}(\tau)$ is rather sensitive to the value of $\alpha$. This will allow us to determine if the system can exhibit quantum behavior even though the known nonclassicality conditions given by both Equations (21) and (22) for the coherence function $g^{(2)}(\tau)$ are violated or, conversely, if the system exhibits nonclassical behavior even though the nonclassicality criterion (20) for $P(\alpha)$ is violated.

It is interesting that Equation (20) for the nonclassicality of $P(\beta)$ is independent of the coherent parameter $\alpha$ since the eigenvalues of the quadratic form $(a^2f^2+b^2d^2+cfd)$ in the exponential in (17) are independent of $\alpha$ while the coherence function $g^{(2)}(\tau)$ is rather sensitive to the value of $\alpha$. The validity of any one of the inequalities in Equations (21) and (22) is sufficient but none of them is actually necessary for nonclassicality. On the other hand, the nonclassicality criterion (20) for the one-time function $P(\alpha)$ may not determine the nonclassicality of the two-time correlation function $g^{(2)}(\tau)$ and conversely. Therefore, condition (20) cannot be a necessary and sufficient condition for nonclassicality since when violated, implying thereby the system is in a classical state, nonetheless the two-time correlation function exhibits nonclassical behavior. The numerical results for $g^{(2)}(\tau)$, as given by Figures 5 and 6, attest to this conclusion, where (20) gives classical behavior from condition (20) for $P(\beta)$ for $\Omega \tau \leq 0.4493$, since $(2\bar{n}+1)e^{-2(\Omega \tau +r)}= 2.4562 e^{-2\Omega \tau}\geq 1$ for $\Omega\tau\leq0.4493$, whereas, both Figures 5 and 6 indicate nonclassical behavior for $0<\Omega \tau<0.5605$.  To minimize intensity fluctuations, it is always optimal to squeeze the amplitude quadrature, that is, to choose $\theta= 2\varphi$, which we impose on all our numerical work.

Figures 1 and 2 show the strictly classical features of the correlation function $g^{(2)}(\tau)$ for $n=0.1$, $r=0.1$, and $|\alpha|=0$ since $g^{(2)}(\tau)$ violates the nonclassical inequalities given by Equations (21) and (22). Note, however, that the nonclassical inequality (20) is satisfied for $\Omega \tau\geq 0$ since $(2\bar{n} +1)e^{-2r}=0.9825<1$. Accordingly, a quasiprobability distribution $P(\beta)$ does not exist since $P(\beta)$ does not vanish as $|\beta|\rightarrow \infty$ nonetheless the correlation $g^{(2)}(\tau)$ exhibits classical behavior. Thus the nonclassical nature of the radiation field, according to the $P(\beta)$ criteria, does not imply that the correlation $g^{(2)}(\tau)$ must behave nonclassically.

In order to show the strong dependence of the coherence function $g^{(2)}(\tau)$ on the coherent parameter $\alpha$, we show in Figure 3 the behavior of $g^{(2)}(\tau)$ for the same values $\bar{n}=0.1$ and $r=0.1$ as those in Figures 1 but with the value of $|\alpha| =2$. In Figure 3, both nonclassical inequalities in (21) are satisfied. In Figure 4, we plot the variable associated with inequality (22) that shows classicality for $0\leq \Omega\tau\leq 2.5793$ and nonclassicality for $\Omega\tau> 2.5793$. Thus the nonclassical nature of the radiation field, according to the $P(\beta)$ criteria (20), can give rise also to mixed classical/nonclassical behavior in the correlation $g^{(2)}(\tau)$ .

The $\lim _{\tau\rightarrow \infty} (g^{(2)}(\tau)-g^{(2)}(0))=0$ gives the critical value of $|\alpha|$, for given $\bar{n}$ and $r$, for which the inequality sign of the second inequality in (21) changes direction. That is, a critical point from classicality to nonclassicality. For instance, for the cases in Figures 1 and 3, $\bar{n}=0.1$, $r=0.1$, the critical value is $|\alpha_{c}|= 0.45397$. That is, $g^{(2)}(\infty)> g^{(2)}(0)$ for $|\alpha|>0.45397$ and  $g^{(2)}(\infty)<g^{(2)}(0)$ for $|\alpha|<0.45397$.

Figures 5 and 6 show the mixed classical/nonclassical nature of both $g^{(2)}(\tau)$ and $(|g^{(2)}(0)-1|-|g^{(2)}(\tau)-1|)$ for $\bar{n}=1.0$, $r=0.1$, and $|\alpha|=0$. In view of inequalities (21) and (22), both functions have nonclassical behavior for $0<\Omega \tau < 0.5605$ and classical for $\Omega \tau \geq 0.5605$. The nonclassicality criterion (20) indicates that a quasiprobability distribution $P(\beta)$ exhibits classical behavior for $0\leq\Omega \tau\leq 0.4493$ and nonclassical for $\Omega \tau> 0.4493$. Therefore, studies of the temporal second-order quantum mechanical correlation function $g^{(2)}(\tau)$, for instance, using a Hanbury Brown-Twiss intensity interferometer modified for homodyne detection \cite{GSSRL07}, will show the nonclassical nature of the correlation. This is contrary to what the nonclassicality criterion (20) would indicate. One must recall that the difference between criterion (20) and criteria (21) and (22) is that the former is based on one-time measurement or behavior of the system whereas the latter involves two-time measurements.

Finally, Figure 7 shows the Mandel $Q_{M}(\tau)$ parameter for $\bar{n} =0.1$, $r=0.1$, and $|\alpha|=2$. The system exhibits nonclassical behavior for $0\leq \Omega \tau <1.7704$ and classical for $\Omega \tau \geq 1.7704$. The field is photon-number squeezed and exhibits sub-Poissonian statistics since $-1\leq Q_{M}(\tau) < 0$. Notice from Figures 3, 4, and 7 that nonclassical effects often occur together but each can occur in the absence of the others.

\begin{figure}
\begin{center}
   \includegraphics[scale=0.3]{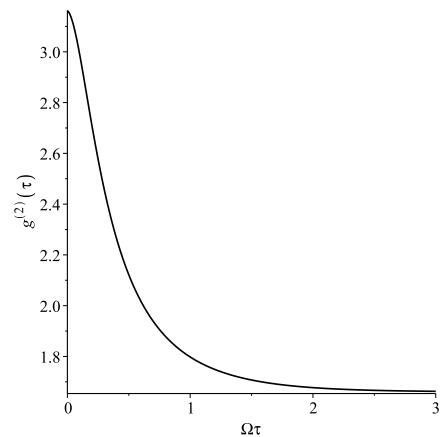}
\end{center}
\label{fig:theFig}
  \caption{Temporal second-order correlation function $g^{(2)}(\tau)$ for $\bar{n}=0.1$, $r=0.1$, and $|\alpha|=0$. One has $g^{(2)}(0) = 3.1625$ and $\lim_{\tau\rightarrow \infty } g^{(2)}(\tau)= 1.6603$. Both nonclassical inequalities in (21) are violated and the statistics is super-Poissonian.}
\end{figure}

\begin{figure}
\begin{center}
   \includegraphics[scale=0.3]{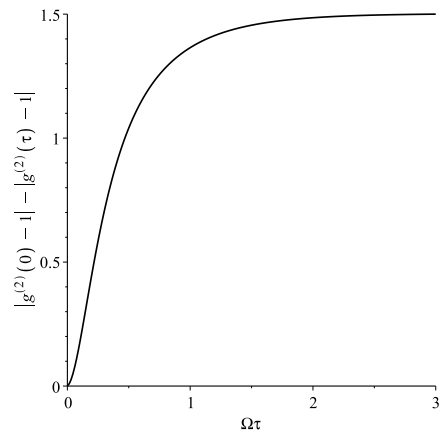}
\end{center}
\label{fig:theFig}
  \caption{Plot of $(|g^{(2)}(0)-1|-|g^{(2)}(\tau)-1|)$ for $\bar{n}=0.1$, $r=0.1$, and $|\alpha|=0$, which asymptotically approaches 1.5022. The nonclassical inequality (22) is violated and $g^{(2)}(\tau)$ is strictly classical.}
\end{figure}

\begin{figure}
\begin{center}
   \includegraphics[scale=0.3]{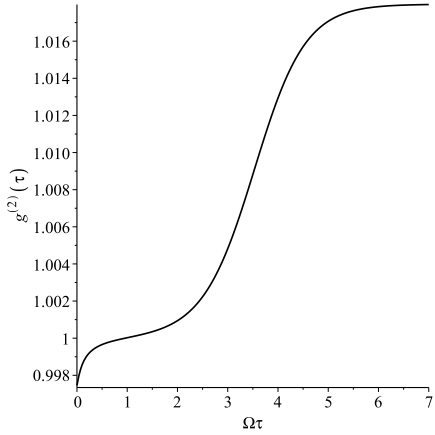}
\end{center}
\label{fig:theFig}
  \caption{Temporal second-order correlation function $g^{(2)}(\tau)$ for $\bar{n}=0.1$, $r=0.1$, and $|\alpha|=2$. One has $g^{(2)}(0) = 0.9975$ and $\lim_{\tau\rightarrow \infty } g^{(2)}(\tau)= 1.0180$. The correlation is strictly nonclassical since both inequalities in (21) are satisfied. The statistics is sub-Poissonian.}
\end{figure}

\begin{figure}
\begin{center}
   \includegraphics[scale=0.3]{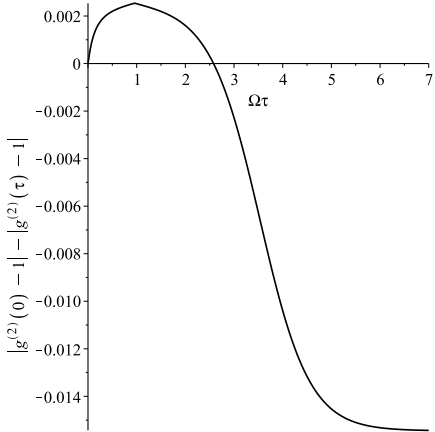}
\end{center}
\label{fig:theFig}
  \caption{Plot of $(|g^{(2)}(0)-1|-|g^{(2)}(\tau)-1|)$ for $\bar{n}=0.1$, $r=0.1$, and $|\alpha|=2$, which asymptotically approaches -0.0155. The behavior is classical for $\Omega \tau\leq 2.5793$ and nonclassical for $\Omega \tau> 2.5793$ according to (22).}
\end{figure}

\begin{figure}
\begin{center}
   \includegraphics[scale=0.3]{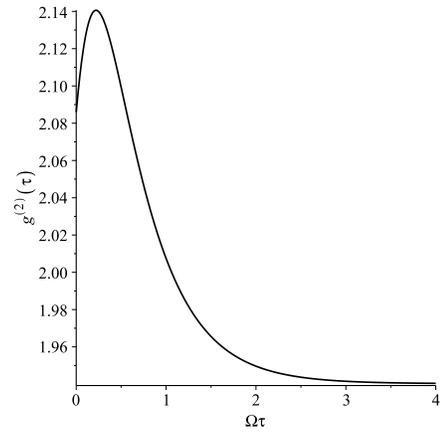}
\end{center}
\label{fig:theFig}
  \caption{Temporal second-order correlation function $g^{(2)}(\tau)$ for $\bar{n}=1.0$, $r=0.1$, and $|\alpha|=0$, where $g^{(2)}(0) = 2.0859$ and $\lim_{\tau\rightarrow \infty } g^{(2)}(\tau)= 1.9402$. The behavior is nonclassical for $0<\Omega \tau<0.5605$ and classical for $\Omega\tau \geq 0.5605$.}
\end{figure}

\begin{figure}
\begin{center}
   \includegraphics[scale=0.3]{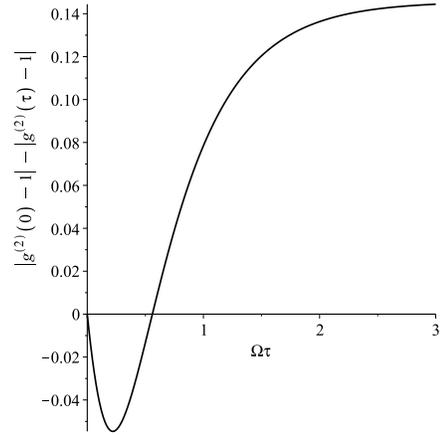}
\end{center}
\label{fig:theFig}
  \caption{Plot of $(|g^{(2)}(0)-1|-|g^{(2)}(\tau)-1|)$ for $\bar{n}=1.0$, $r=0.1$, and $|\alpha|=0$, which asymptotically approaches 0.1457. The behavior is nonclassical for $0<\Omega \tau <0.5605$ and classical for $\Omega \tau\geq 0.5605$.}
\end{figure}

\begin{figure}
\begin{center}
   \includegraphics[scale=0.3]{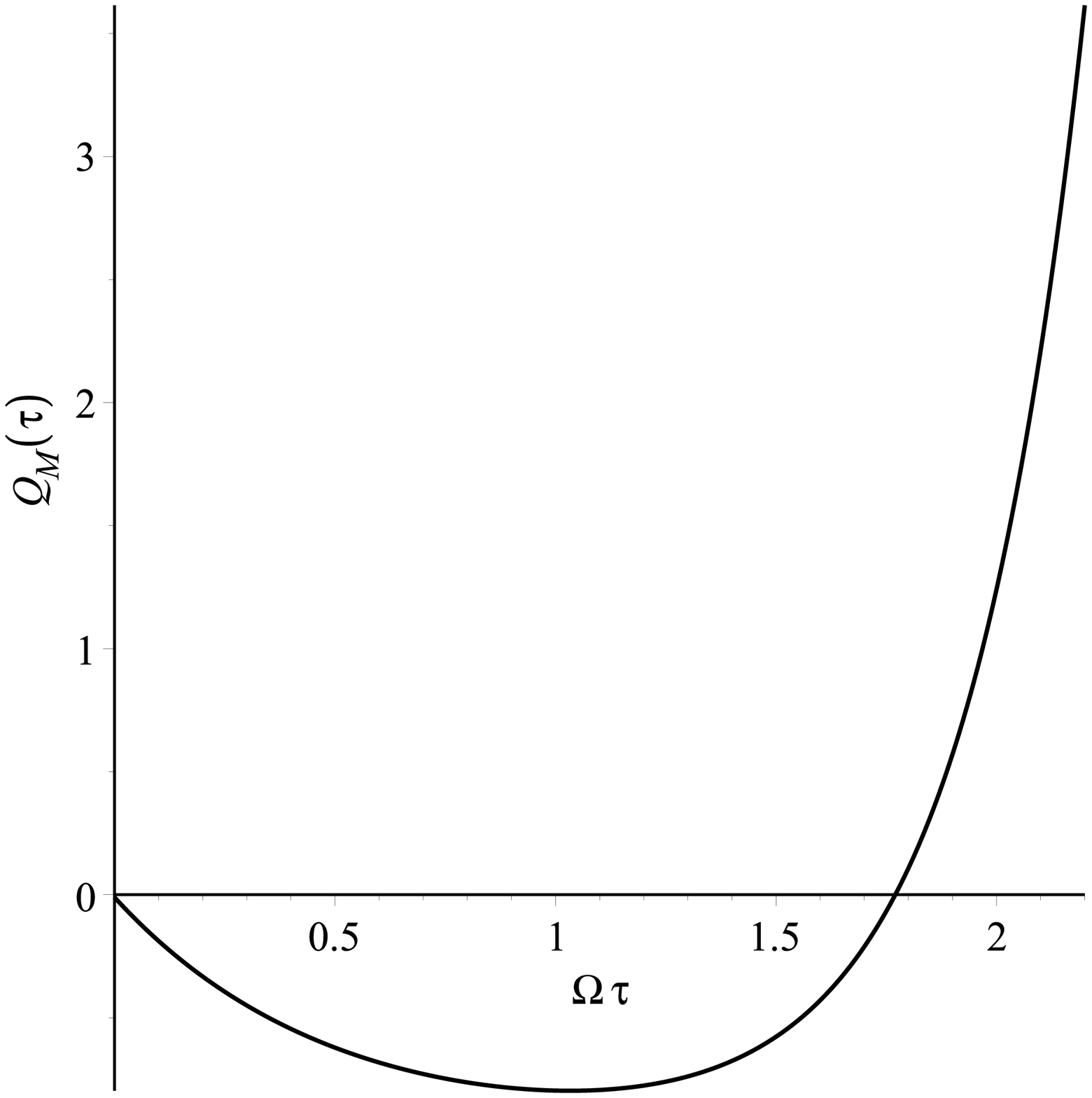}
\end{center}
\label{fig:theFig}
  \caption{Plot of the Mandel parameter $Q_{M}(\tau)$ for $\bar{n}=0.1$, $r=0.1$, and $|\alpha|=2$. The behavior is nonclassical for $0\leq \Omega \tau < 1.7704$ and classical for $\Omega \tau\geq 1.7704$ since $Q_{M}(0)=-0.0104$.}
\end{figure}

\section{Summary and discussions}

We calculate the one-time quasiprobability distribution $P(\beta)$ and the two-time, second-order coherence function $g^{(2)}(\tau)$ for Gaussian states (2), viz., displaced-squeezed thermal states, where the dynamics is governed solely by the general, degenerate parametric amplification Hamiltonian (1). We use these exact results to analyze the different characterization of nonclassicality. We find from our numerical studies that satisfying any of the conditions for the coherence function $g^{(2)}(\tau)$ given in Equations (21) and (22) are sufficient for nonclassicality; however, violations of both conditions (21) and (22) does not insure strictly classical behavior. We find examples whereby the nonclassicality condition (20) for $P(\beta)$ is satisfied while the coherence function $g^{(2)}(\tau)$ satisfies all the known classical conditions and conversely, whereby the nonclassicality condition (20) is violated, that is, the quasiprobability distribution $P(\beta)$ exists, nonetheless, the coherence function $g^{(2)}(\tau)$ exhibits nonclassical behavior. Therefore, it does not seem possible to find a single set of necessary and sufficient conditions, based on the state of the system and measurements of observables of the system, which would unequivocally establish the classical or nonclassical nature of the radiation field.

\appendix*
\section{Second-order coherence}
The degree of second-order temporal coherence is \cite{MA16}
\begin{equation}
g^{(2)}(\tau)= 1 + \frac{n^2(\tau)+ s^2(\tau) +u(\tau) n(\tau) -v(\tau)s(\tau)}{\langle \hat{a}^{\dag}(0)\hat{a}(0)\rangle \langle \hat{a}^{\dag}(\tau)\hat{a}(\tau) \rangle},
\end{equation}
where
\begin{equation}
n(\tau)= (\bar{n}+ 1/2)\cosh \big{(}  \Omega\tau + 2r\big{)} -(1/2)\cosh (\Omega\tau),
\end{equation}

\begin{equation}
s(\tau)= (\bar{n}+ 1/2)\sinh \big{(}\Omega\tau + 2r\big{)} -(1/2)\sinh (\Omega\tau),
\end{equation}

\begin{equation}
u(\tau) = \alpha A^*(\tau) + \alpha^* A(\tau),
\end{equation}
and
\begin{equation}
v(\tau)=\alpha A(\tau)\exp{(-i\theta)}+\alpha^* A^*(\tau)\exp{(i\theta)},
\end{equation}
where $A(\tau)$ is defined by Equation (10).

The time development of the photon number is given by
\[
\textup{Tr}[\hat{\rho}(t+\tau)\hat{a}^\dag \hat{a}] = \langle \hat{a}^\dag(\tau) \hat{a}(\tau)\rangle = \langle\hat{n}(\tau)\rangle
\]
\begin{equation}
=(\bar{n}+1/2)\cosh[2(\Omega\tau+r)] + |A(\tau)|^2 -\frac{1}{2}.
\end{equation}

Equation (10) is the correct expression for $A(\tau)$ and not that given in Ref. \cite{MA16}, where in Equation (13) the purely imaginary number $i$ should not be there. Similarly, there is no $i$ in the square braces of  Equation (A2) in Ref. \cite{MA16}.

\begin{newpage}
\bibliography{}

\end{newpage}
\end{document}